\documentstyle[amsmath,epsf,a4,11pt]{article}
\begin{document}

\date{}
\title{{\bf Dynamics and stability of the G\"{o}del universe}}
\author{John D. Barrow$^1$\thanks{e-mail address:
j.d.barrow@damtp.cam.ac.uk} and Christos G.
Tsagas$^{1,2}$\thanks{e-mail address: c.tsagas@damtp.cam.ac.uk}\\
{\small $^1$DAMTP, Centre for Mathematical Sciences, University of
Cambridge,}\\ {\small Cambridge~CB3~0WA, UK}\\{\small
$^2$Department of Mathematics and Applied Mathematics, University
of Cape Town,}\\ {\small Rondebosch 7701, South Africa}}

\maketitle

\begin{abstract}
We use covariant techniques to describe the properties of the
G\"{o}del universe and then consider its linear response to a
variety of perturbations. Against matter aggregations, we find
that the stability of the G\"{o}del model depends primarily upon
the presence of gradients in the centrifugal energy, and
secondarily on the equation of state of the fluid. The latter
dictates the behaviour of the model when dealing with homogeneous
perturbations. The vorticity of the perturbed G\"{o}del model is
found to evolve as in almost-FRW spacetimes, with some additional
directional effects due to shape distortions. We also consider
gravitational-wave perturbations by investigating the evolution of
the magnetic Weyl component. This tensor obeys a simple plane-wave
equation, which argues for the neutral stability of the G\"{o}del
model against linear gravity-wave distortions. The implications of
the background rotation for scalar-field G\"{o}del cosmologies are
also discussed.
\end{abstract}

\section{Introduction}
The G\"{o}del universe is an exact solution of the Einstein field
equations, which is both stationary and spatially
homogeneous~\cite{G}. The model, which is of Petrov type $D$, is
also rotationally symmetric about each point and contains a
perfect-fluid matter source whose 4-velocity is a Killing
vector~\cite{RS}. G\"{o}del's universe is known for its unusual
global properties. The most intriguing among them is the existence
of closed timelike curves, which violate global causality and
makes time-travel theoretically possible in this space-time.
Hermann Weyl had first suggested in 1921 that time travel might
occur in general relativity \cite{weyl} as a consequence of "very
considerable fluctuations in the space-time metric" but he
believed that the fluctuations "necessary to produce this effect
do not occur in the region of the world in which we live". It was
clear that this causal anomaly could arise for {\it some}
distorted geometry and hence for some distribution of mass and
energy. Kurt G\"{o}del showed that the required distribution could
be extremely simple and his discovery initiated the study of the
global properties and causal structure of Einstein's equations
\cite{HE}. At first, attention was almost entirely focussed upon
the properties of the equations of motion in the G\"{o}del
universe \cite{kun}-\cite{stein} and an interesting overview of
these investigations has recently been given by Ozsv\'{a}th and
Sch\"{u}cking~\cite{osz}. Subsequently, G\"{o}del's solution has
also triggered a considerable amount of work on rotating solutions
of the Einstein field equations and rotational cosmological models
(see~\cite{NST}-\cite{Ca} for a representative list). For further
details and an extensive discussion the reader is referred to the
recent review articles by Krasi\'{n}ski~\cite{K1}-\cite{K3} and
Obukhov~\cite{O}. Although the G\"{o}del spacetime is not a
realistic model for our universe, it is an important theoretical
laboratory for investigating a range of global properties of the
spacetime structure in different gravity theories. Recent studies
of quantum computation~\cite{deutsch, bacon} have shown that the
presence of closed timelike curves in spacetime provides for a new
physical model for quantum computation in compact regions. A
quantum computer with access to closed timelike curves can solve
NP-complete problems with only a polynomial number of quantum
logic gates. A series of studies of the possibility of closed
timelike paths in string theories~\cite{BD}-\cite{har} have also
drawn on the insights gained from the detailed study of the
G\"{o}del spacetime and its generalisations to arbitrary space
dimensions \cite{gurses, brech1}, considering the constraints of
possible holographic principles~\cite{brecher}, and investigating
the accessibility of information in different parts of
spacetime~\cite{hik}.

The G\"{o}del universe is a member of the family of homogeneous
spacetimes together with the Einstein static and de Sitter
universes. In recent years close attention has been paid to the
stability of the de Sitter universe and its physical significance
for the consequences of inflation in the early stages of our
universe. Under physically realistic conditions on the
energy-momentum tensor of matter, all perturbations to isotropy
and homogeneity will be seen to fall off exponentially rapidly
within the event horizon of a geodesically moving observer in the
de Sitter universe~\cite{JDB}-\cite{W}. This result, in its
different technical expressions, is known as the cosmic no hair
theorem. The stability of the Einstein static universe has also
been examined in a covariant fashion and reveals a more intricate
dependence on material content than is conventionally reported
\cite{Gib, BEMT}. We will complement these detailed studies of the
stability of the de Sitter and Einstein static homogeneous
spacetimes with a covariant study of the stability of the
G\"{o}del universe. Despite the past interest in the causal
structure of the G\"{o}del model there have been few perturbative
studies of its properties and these have been confined to the
investigation of the effects of the global rotation on the
evolution of scalar density perturbations in the papers of
Silk~\cite{S1,S2}. Here, we investigate the stability with respect
to scalar, vector and tensor perturbation modes using the gauge
covariant formalism of Ellis and Bruni~\cite{EB}. In the process
we show that the background vortical energy contributes to the
gravitational pull of the matter, while it gradients add to the
pressure support. The balance between these two agents effectively
determines the stability of the G\"odel universe against matter
aggregations. By examining the rotational behaviour of the
perturbed model, we explain how shape distortions can increase or
decrease its overall rotation. We also introduce a set of linear
covariant constraints, which isolate the pure tensor
perturbations, and analyse their propagation on the rotating
G\"{o}del background. Our results argue for the neutral stability
of these gravitational-wave distortions. Finally, we consider some
of the effects of spatially homogeneous perturbations and discuss
the compatibility of scalar fields with the symmetries of the
G\"{o}del spacetime.

\section{Covariant characterisation of the G\"odel universe}
\subsection{The irreducible kinematical variables}
The covariant description of the G\"{o}del universe, with respect
to its irreducible kinematical quantities, has been given
in~\cite{E1} (see also~\cite{HE,E2}). Here, we will briefly
introduce and extent this description and also present the
associated constraints.

Relative to a timelike 4-velocity field $u_{a}$ (normalised so
that $u_{a}u^{a}=-1$) that is tangent to the worldlines of the
fundamental observers, the G\"{o}del spacetime is covariantly
described by~\cite{E1}
\begin{equation}
\Theta=0=\dot{u}_{a}=\sigma_{ab}\,, \hspace{10mm} {\rm and}
\hspace{10mm} \omega_{a}\neq 0\,.  \label{Gkin}
\end{equation}
Therefore, with the exception of the vorticity ($\omega_{a}$), the
rest of the kinematical variables, namely the volume expansion
($\Theta$), the shear ($\sigma_{ab}$) and the acceleration
($\dot{u}_{a}$), vanish identically. We note that
$\nabla_{b}\omega_{a}=0$, so ensuring that the vorticity vector
associated with the 4-velocity field $u_{a}$ is covariantly
constant.

\subsection{The twice-contracted Bianchi identities}
The stationary nature of the G\"{o}del solution means that there
are no propagation equations: they have all been transformed into
constraints. For example, on using the twice-contracted Bianchi
identities we obtain the standard conservation laws for the energy
and momentum densities. When applied to the G\"{o}del spacetime,
the latter yield the constraints
\begin{equation}
\dot{\mu}=0\hspace{10mm}{\rm and}\hspace{10mm}{\rm D}_{a}p=0\,,
\label{conslaws}
\end{equation}
where $\mu$ and $p$ are the matter density and isotropic pressure
respectively.\footnote{Originally, G\"{o}del's solution was given
for dust (i.e.~$\mu\neq0$, $p=0$, $\Lambda\neq0$). However,
through the transformation $\mu\rightarrow\mu^{\prime}=\mu+p$ and
$\Lambda\rightarrow\Lambda^{\prime}=\Lambda+\kappa p$, the
G\"{o}del spacetime can be reinterpreted as a perfect-fluid
model.} Here, an overdot indicates differentiation along $u_{a}$
(e.g.~$\dot{\mu}=u^{a}\nabla_{a}\mu$). Also, ${\rm
D}_{a}=h_{a}{}^{b}\nabla_{b}$ is the covariant derivative operator
orthogonal to $u_{a}$, and $h_{ab}=g_{ab}+u_{a}u_{b}$ is the
associated projection tensor.

\subsection{The Ricci identities}
Covariantly, the kinematic evolution is determined by a set of
three propagation equations and three constraints, all of which
derive from the Ricci identities $\nabla_{\lbrack
a}\nabla_{b]}u_{c}=R_{dcba}u^{d}$, where $R_{abcd}$ is the
spacetime Riemann tensor. In G\"{o}del's universe, the
Raychaudhuri equation, which describes the volume evolution of a
fluid element, reduces to
\begin{equation}
{\textstyle{\frac{1}{2}}}\kappa(\mu +3p)- 2\omega^{2}-
\Lambda=0\,,  \label{Ray}
\end{equation}
with $\kappa=8\pi G$ and $\omega^{2}=\omega_{a}\omega^{a}$. Note
how the rotation balances the gravitational attraction of the
matter, as well as that of the (negative - see Sec.~2.5)
cosmological constant. Thus, in the G\"{o}del model the vorticity
has assumed the role played by the (positive) cosmological
constant in the Einstein static universe. Of the two remaining
propagation equations, the shear evolution formula takes the form
\begin{equation}
E_{ab}+ \omega_{\langle a}\omega_{b\rangle}=0\,,  \label{sigmadot}
\end{equation}
where $E_{ab}$ is the electric component of the Weyl
tensor.\footnote{Angled brackets are used to indicate orthogonally
projected vectors and the projected, symmetric and trace-free part
of second rank tensors.} The vorticity propagation equation, on
the other hand, is trivially satisfied. Two of the three
kinematical constraints, the shear-divergence and the
vorticity-divergence, are also trivially satisfied, while the
gravito-magnetic constraint leads to
\begin{equation}
H_{ab}=0\,,  \label{Hab}
\end{equation}
with $H_{ab}$ being the magnetic counterpart of $E_{ab}$. It
should be emphasised that the latter is not covariantly constant
(i.e.~$\nabla_{c}E_{ab}\neq0$). Instead, one can use
Eq.~(\ref{sigmadot}) to show that
\begin{equation}
\dot{E}_{ab}=0={\rm D}_{c}E_{ab}\,,  \label{Eab}
\end{equation}
in agreement with the stationary nature and spatial homogeneity of
the G\"{o}del spacetime. Clearly, result (\ref{Eab}b) also
guarantees that ${\rm D}^{b}E_{ab}=0={\rm curl}E_{ab}$.

\subsection{The Bianchi identities}
The Bianchi identities provide two pairs of propagation and
constraint equations for the conformal curvature, which is
monitored through the electric and the magnetic parts of the Weyl
tensor. When applied to the G\"{o}del spacetime, the
$\dot{E}$-equation gives
\begin{equation}
E_{\langle a}{}^{d}\epsilon_{b\rangle cd}\omega^{c}=0\,,
\label{dotE}
\end{equation}
while the $\dot{H}$-equation is trivially satisfied. Note that,
when (\ref{sigmadot}) is taken into account, the above constraint
also becomes trivial. On the other hand, the ${\rm div}E_{ab}$ and
${\rm div}H_{ab}$ constraints associated with the Bianchi
identities lead to
\begin{equation}
{\rm D}_{a}\mu=0  \label{Econ}
\end{equation}
and
\begin{equation}
\kappa(\mu +p)\omega _{a}+ 3E_{ab}\omega^{b}=0\,,  \label{Hcon}
\end{equation}
respectively. Results (\ref{conslaws}a) and (\ref{Econ}) combine
to ensure that $\nabla_{a}\mu=0$, while from Eq.~(\ref{Ray}) we
obtain $\dot{p}=0$. In other words, both the energy density and
the isotropic pressure of the matter that fill G\"{o}del's
universe are covariantly constant quantities.

\subsection{Further constraints}
Additional constraints are obtained by contracting
(\ref{sigmadot}) along $\omega_{a}$ and substituting the result
into Eq.~(\ref{Hcon}). Then one finds that
\begin{equation}
\kappa(\mu+p)=2\omega^{2}\,,  \label{rho+p}
\end{equation}
which also measures the total inertial mass of the G\"{o}del
universe. On using this result, we can recast Eq.~(\ref{Ray}) as
\begin{equation}
\kappa(\mu-p)=-2\Lambda\,,  \label{rho-p}
\end{equation}
to guarantee that $\Lambda\leq0$ as long as $p\leq\mu$. The
cosmological constant vanishes only when the fluid has a maximally
stiff equation of state, with $p=\mu$. Thus, for conventional
matter sources, the G\"{o}del spacetime has non-positive $\Lambda
$. This restriction is relaxed when dealing with the Newtonian
analogue of the G\"{o}del universe (see below). From (\ref{rho+p})
and (\ref{rho-p}) it becomes clear that, given the equation of
state of the matter, only one of $\mu$, $\omega$ or $\Lambda$ is
needed to determine the other two. Finally, constraints
(\ref{rho+p}) and (\ref{rho-p}) combine to give
\begin{equation}
\kappa\mu+ \Lambda- \omega^{2}=0\,,  \label{r+L-O2}
\end{equation}
which is the G\"{o}del analogue of the Friedmann equation.

The generic rotation of the G\"{o}del universe means that the
fluid flow lines are not hypersurface orthogonal and, therefore,
there are no integrable spatial sections. Nevertheless, one can
still employ the generalised Gauss-Codacci equation to evaluate
the orthogonally projected Ricci tensor. Applied to the G\"{o}del
model, the latter reads
\begin{equation}
{\cal R}_{ab}=
{\textstyle{\frac{2}{3}}}(\kappa\mu+\Lambda-\omega^{2})h_{ab}\,,
\label{cR<ab>}
\end{equation}
which by means of constraint (\ref{r+L-O2}) ensures that ${\cal
R}_{ab}$ vanishes.

\subsection{The Newtonian analogue}
There is a simple Newtonian counterpart to the rotating G\"{o}del
universe in the case of zero pressure that has been explored by
Ozsv\'{a}th and Sch\"{u}cking~\cite{OS1} (see also~\cite{HS})
following G\"{o}del~\cite{God}. If we consider a pressureless
fluid of constant density $\mu$ rotating rigidly with constant
angular velocity $\omega$ about the $z-$axis so $\vec{v}=(-\omega
y,\omega x,0)$ then, by integrating the Euler equation
\begin{equation}
\frac{\partial \vec{v}}{\partial t}+
\vec{v}\cdot\nabla\vec{v}=-\nabla\Phi, \label{eul}
\end{equation}
we find that the Newtonian gravitational potential is
\begin{equation}
\Phi={\textstyle{1\over2}}\omega^{2}(x^{2}+y^{2})\,.  \label{pot}
\end{equation}
where now $\nabla$ indicates ordinary partial derivatives. This
solves the Poisson equation with cosmological constant term
($c=1$),
\begin{equation}
\nabla^{2}\Phi+ \Lambda={\textstyle{1\over2}}\kappa\mu\,,
\label{poi}
\end{equation}
if
\begin{equation}
2\omega^{2}\equiv{\textstyle{1\over2}}\kappa\mu- \Lambda\,.
\label{newt}
\end{equation}
We see that this Newtonian result coincides with the relation
(\ref{Ray}) for the G\"{o}del solution in the \ $p=0$ case and, in
contrast to general relativity, can be satisfied even for
$\Lambda=0$ and some $\Lambda >0$ values. In general relativity
the additional geometrical constraint (\ref{rho-p}) leads to the
stronger relation,
\begin{equation}
\omega^{2}\equiv{\textstyle{1\over2}}\kappa\mu\,,  \label{g}
\end{equation}%
for G\"{o}del's solution, which therefore requires $\Lambda<0$.

\subsection{The closed timelike curves}
The most intriguing property of the G\"{o}del spacetime is the
existence of closed timelike curves, which makes time travel a
theoretical possibility and has ensured an enduring interest in
the model~\cite{RT}-\cite{OS2}. To demonstrate the presence of
such closed timelike worldlines, consider the line element of the
original G\"{o}del solution (i.e.~with $p=0$ and metric signature
-2) written in cylindrical coordinates ($r$, $z$, $\phi$)
\begin{equation}
{\rm d}s^{2}=4a^{2}\left[{\rm d}t^{2}-{\rm d}r^{2}-{\rm d}z^{2}
+\left(\sinh^{4}r-\sinh^{2}r\right){\rm d}\phi^{2}
+2\sqrt{2}\sinh^{2}r{\rm d}\phi{\rm d}t\right]\,, \label{Godel}
\end{equation}
with $1/a^{2}=\kappa\mu=-2\Lambda$~\cite{G}. Then, introducing the
coordinate transformation $t\rightarrow\tilde{t}=2at$,
$r\rightarrow\tilde{r}=2ar$, $z\rightarrow\tilde{z}=2az$ and
$\phi\rightarrow\tilde{\phi}=\phi$, transform the above line
element into
\begin{equation}
{\rm d}s^{2}={\rm d}t^{2}-{\rm d}r^{2}-{\rm d}z^{2}+
4a^{2}\left[\sinh^{4}\left({\frac{r}{2a}}\right)
-\sinh^{2}\left({\frac{r}{2a}}\right)\right]{\rm d}\phi^{2}+
4\sqrt{2}a\sinh^{2}\left({\frac{r}{2a}}\right){\rm d}\phi{\rm
d}t\,,  \label{Godel1}
\end{equation}
where the tildas have been suppressed. Clearly, if for some
$r=r_{0}$ the quantity $\sinh^{4}(r/2a)-\sinh^{2}(r/2a)$ is
positive, the circle defined by $r=r_{0}$, $t=0=z$ will be
timelike everywhere~\cite{G}. Thus, on using (\ref{rho+p}) with
$p=0$, the condition $r>2\ln(1+\sqrt{2})/\sqrt{\kappa\mu}$ for the
existence of such closed timelike curves reads
\begin{equation}
r>R_{G}\equiv\frac{\sqrt{2}\ln(1+\sqrt{2})}{\omega }\,,
\label{tlcond}
\end{equation}
where $R_{G}$ denotes the radius of the observer's
\textquotedblleft causal region\textquotedblright . Hence,
$R_{G}\rightarrow \infty $ as $\omega \rightarrow 0$, which means
that the weaker the rotation of the model the more
\textquotedblleft remote\textquotedblright\ the closed timelike
curves become. Alternatively, one might say that the faster the
G\"{o}del universe rotates the smaller its causal region becomes.
Note that one can include the fluid pressure in Eq.~(\ref{tlcond})
by simply using the transformation
$\mu\rightarrow\tilde{\mu}=\mu+p$.

\section{The perturbed G\"odel universe}
\subsection{Conservation laws}
For a perfect fluid, the energy density and the momentum density
conservation laws are given by the nonlinear expressions
\begin{equation}
\dot{\mu}=-(\mu +p)\Theta \,,  \label{edc}
\end{equation}%
and
\begin{equation}
(\mu+p)\dot{u}_{a}=-{\rm D}_{a}p\,,  \label{mdc}
\end{equation}
respectively. These formulae also hold after we have linearised
about the G\"{o}del background. In that case, however, we can
substitute the zero-order relation $\mu+p=2\omega^{2}$ on the
right-hand side of (\ref{edc}). Also, when dealing with a
barotropic perfect fluid (i.e.~for $p=p(\mu)$), we have ${\rm
D}_{a}p=c_{{\rm s}}^{2}{\rm D}_{a}\mu $, where $c_{{\rm
s}}^{2}={\rm d}p/{\rm d}\mu $ is the square of the adiabatic sound
speed.

\subsection{Density perturbations}
Consider a general spacetime filled with a single barotropic
perfect fluid. The nonlinear evolution of density inhomogeneities
is monitored by the system~\cite{EB}
\begin{eqnarray}
\dot{{\cal D}}_{\langle a\rangle}&=&w\Theta{\cal D}_{a}-
(\sigma_{ab}-\omega_{ab}){\cal D}^{b}-(1+w){\cal Z}_{a}\,,
\label{nlcDdot}\\ \dot{{\cal Z}}_{\langle a\rangle}&=&
-{\textstyle{\frac{2}{3}}}\Theta{\cal Z}_{a}-
(\sigma_{ab}-\omega_{ab}){\cal Z}^{b}-
{\textstyle{\frac{1}{2}}}\kappa\mu{\cal D}_{a}+ a\Re\dot{u}_{a}+
aA_{a}- 2a{\rm D}_{a}(\sigma^{2}-\omega^{2})\,,  \label{nlcZdot}
\end{eqnarray}
where ${\cal D}_{a}=(a/\mu){\rm D}_{a}\mu$, ${\cal Z}_{a}=a{\rm
D}_{a}\Theta$ and $w\equiv p/\mu $ by definition.\footnote{The
background scale $a$, which has been introduced to make ${\cal D}$
and ${\cal Z}$ dimensionless and appears in Eqs.~(\ref{nlcDdot}),
(\ref{nlcZdot}), is a characteristic dimension of the G\"{o}del
universe corresponding to the radius of the smallest closed
timelike curves~\cite{G,S2}.} Also
\begin{equation}
\Re=\kappa\mu- {\textstyle{\frac{1}{3}}}\Theta^{2}-
2(\sigma^{2}-\omega^{2})+ A+ \Lambda\,,  \label{nlRe}
\end{equation}
defines $\Re$ and $A_{a}={\rm D}_{a}A$ with
$A=\nabla^{a}\dot{u}_{a}$. The 4-acceleration satisfies the
momentum-density conservation law, which for adiabatic
perturbations is given by Eq.~(\ref{mdc}). The projected vectors
${\cal D}_{a}$ and ${\cal Z}_{a}$ respectively describe local
inhomogeneities in the matter energy density and in the volume
expansion~\cite{EB}. Note that in an exact G\"{o}del universe
${\rm D}_{a}\mu=0={\rm D}_{a}\Theta $. Hence, ${\cal D}_{a}$ and
${\cal Z}_{a}$ vanish identically to zero order and therefore both
satisfy the gauge-invariant requirements~\cite{SW}.

The scalar $\Re$ plays an important and subtle role via its
coupling with the 4-acceleration in Eq.~(\ref{nlcZdot}). In
particular, the sign of $\Re$ determines the effect of the
pressure gradients associated with $\dot{u}_a$. When $\Re$ is
positive these gradients will add to the gravitational pull of the
density inhomogeneities. Otherwise, they will contribute to the
pressure support. The subtlety in $\Re$ is that the vorticity adds
to the effect of the ordinary matter, whereas the shear opposes
it. This is a purely relativistic effect, with no known Newtonian
analogue. Note that in exact FRW models $\Re=0$, which explains
why the aforementioned behaviour has passed relatively
unnoticed~\cite{EB}.

Linearising Eqs.~(\ref{nlcDdot}), (\ref{nlcZdot}) about the
G\"{o}del universe and keeping only the zero-order component of
(\ref{nlRe}) we obtain
\begin{eqnarray}
\dot{{\cal D}}_{a}&=&\omega _{ab}{\cal D}^{b}-(1+w){\cal Z}_{a}\,,
\label{lcDdot}\\ \dot{{\cal Z}}_{a}&=&\omega_{ab}{\cal Z}^{b}-
{\textstyle{\frac{1}{2}}}\kappa\mu{\cal D}_{a}+ a\Re \dot{u}_{a}+
aA_{a}+ 2a{\rm D}_{a}\omega^{2}\,,  \label{lcZdot}
\end{eqnarray}
with
\begin{equation}
\Re=\kappa\mu+ 2\omega^{2}+ \Lambda=3\omega^{2}>0\,,  \label{lRe}
\end{equation}
since $\kappa\mu-\omega^{2}+\Lambda=0$ in the exact G\"{o}del
spacetime. Also, $A={\rm D}^{a}\dot{u}_{a}$ to first order and
$\dot{u}_{a}$ is still given by Eq.~(\ref{mdc}). Then, the system
(\ref{lcDdot}), (\ref{lcZdot}) reads
\begin{eqnarray}
\dot{{\cal D}}_{a}&=&\omega_{ab}{\cal D}^{b}-(1+w){\cal Z}_{a}\,,
\label{lcDdot1}\\ \dot{{\cal Z}}_{a}&=&\omega_{ab}{\cal Z}^{b}-
{\textstyle{\frac{1}{2}}}\kappa\mu{\cal D}_{a}-
\frac{3a\omega^{2}}{\mu(1+w)}{\rm D}_{a}p- \frac{a}{\mu(1+w)}{\rm
D}_{a}({\rm D}^{2}p)+ 2a{\rm D}_{a}\omega^{2}\,.  \label{lcZdot1}
\end{eqnarray}
where $\omega^{2}=\kappa\mu(1+w)/2$ to zero order (see
Eq.~(\ref{rho+p})). Compared to the perturbed FRW case, the
nonzero rotation of the G\"{o}del background has added four extra
terms to the system (\ref{lcDdot1}), (\ref{lcZdot1}). They are the
first term on the right-hand side of (\ref{lcDdot1}), and the
first, third and the last term on the right-hand side of
(\ref{lcZdot1}). Note the second of the three vorticity terms in
Eq.~(\ref{lcZdot1}), which is proportional to $\omega^{2}$ and
adds to the overall gravitational pull. This relativistic effect
seems to suggest that rotational energy also has \textquotedblleft
weight\textquotedblright. Indirectly, it also seems to favour the
de Felice and the Barrab\`{e}s et al interpretation~\cite{dF,BBI}
of the Abramowicz-Lasota \textquotedblleft centrifugal force
reversal\textquotedblright\ effect~\cite{AL,AP}. The third
vorticity term in (\ref{lcZdot1}), on the other hand, is triggered
by inhomogeneities in $\omega^{2}$ and will also appear in a
Newtonian treatment. It can resist the collapse, thus mimicking
the effects of the ordinary pressure gradients. As we shall see
below, the stability of the G\"{o}del universe depends crucially
on which of these two terms dominates over the other.

Density perturbations are related to gradients in the vorticity
and the curvature by means of the Gauss-Codacci equation,
according to which the linearised projected Ricci scalar is given
by ${\cal R}=2(\kappa\mu-\omega^{2}+\Lambda)$. The above
immediately implies the linear relation
\begin{equation}
\kappa\mu{\cal D}_{a}=a{\rm D}_{a}\omega^{2}+
{\textstyle{\frac{1}{2}}}a {\rm D}_{a}{\cal R}\,,  \label{D2cR}
\end{equation}
between ${\cal D}_a$, ${\rm D}_a\omega^2$ and ${\rm D}_a{\cal R}$.

\subsection{Two alternative types of perturbations}
Following (\ref{D2cR}) we may consider two types of density
perturbation, first by setting ${\rm D}_{a}{\cal R}=0$ and then by
assuming that ${\rm D}_{a}\omega^{2}=0$. For simplicity, we will
label the former {\em isocurvature} and the latter {\em
perturbations under rigid rotation}. In general one does expect
these two perturbation types to propagate independently. However,
this approximation scheme simplifies the system (\ref{lcDdot1}),
(\ref{lcZdot1}) and we can investigate the effects of global
rotation on the evolution of density perturbations analytically.

\subsubsection{Isocurvature perturbations}
Setting ${\rm D}_{a}{\cal R}=0$ in Eq.~(\ref{D2cR}) means that
$a{\rm D}_{a}\omega^{2}=\kappa\mu{\cal D}_{a}$ and the linear
system (\ref{lcDdot1}), (\ref{lcZdot1}) takes the form
\begin{eqnarray}
\dot{{\cal D}}_{a}&=&\omega_{ab}{\cal D}^{b}- (1+w){\cal Z}_{a}\,,
\label{ilcDdot1}\\ \dot{{\cal Z}}_{a}&=&\omega_{ab}{\cal Z}^{b}+
{\textstyle{\frac{3}{2}}}\kappa\mu(1-c_{{\rm s}}^{2}){\cal D}_{a}-
\frac{c_{{\rm s}}^{2}}{1+w}{\rm D}_{a}({\rm D}^{b}{\cal D}_{b})\,,
\label{ilcZdot1}
\end{eqnarray}
where we have used the background relation
$\omega^{2}=\kappa\mu(1+w)/2$ and the barotropic expression ${\rm
D}_{a}p=(\mu c_{{\rm s}}^{2}/a){\cal D}_{a}$. The projected
divergence of the these equations, multiplied by the background
scale $a$, gives the set of equations governing the linear
evolution of isocurvature density perturbations in a perturbed
G\"{o}del universe.
\begin{eqnarray}
\dot{\Delta}&=&-(1+w){\cal Z}\,,  \label{ilDeltadot}\\
\dot{{\cal Z}}&=&{\textstyle{\frac{3}{2}}}\kappa\mu(1-c_{{\rm
s}}^{2})\Delta- \frac{c_{{\rm s}}^{2}}{1+w}{\rm D}^{2}\Delta\,,
\label{ilZetadot}
\end{eqnarray}
where $\Delta=a{\rm D}^{a}{\cal D}_{a}$ and ${\cal Z}=a{\rm
D}^{a}{\cal Z}_{a}$ by definition.\footnote{In deriving
Eq.~(\ref{ilDeltadot}) we have used the linear result $a{\rm
D}^{a}\dot{{\cal D}}_{a}=\dot{\Delta}+a\omega _{ab}{\rm
D}^{a}{\cal D}^{b}$. An exactly analogous relation for the
expansion gradients has been used to obtain (\ref{ilZetadot}).}
Note that the scalar $\Delta $ describes local matter aggregations
and it is the covariant analogue of the standard density contrast
$\delta\rho/\rho $. The system (\ref{ilDeltadot}),
(\ref{ilZetadot}) leads to the following wavelike equation for the
evolution of matter aggregations
\begin{equation}
\ddot{\Delta}=-{\textstyle{\frac{3}{2}}}\kappa\mu(1+w)(1-c_{{\rm
s}}^{2})\Delta+ c_{{\rm s}}^{2}{\rm D}^{2}\Delta\,,
\label{ilDeltaddot}
\end{equation}
and subsequently leads to
\begin{equation}
\ddot{\Delta}_{(k)}=-\left[ {\textstyle{\frac{3}{2}}}\kappa \mu
(1+w)(1-c_{{\rm s}}^{2})+\frac{k^{2}c_{{\rm s}}^{2}}{a^{2}}\right]
\Delta_{(k)}\,, \label{hilDeltaddot}
\end{equation}
assuming the decomposition $\Delta=\sum_{k}\Delta_{(k)}Q^{(k)}$
with ${\rm D}_{a}\Delta_{(k)}=0$. Note that $Q^{(k)}$ are locally
defined scalar harmonics with $\dot{Q}^{(k)}=0$ and ${\rm
D}^{2}Q^{(k)}=-(k^{2}/a^{2})Q^{(k)}$. The last equation has an
oscillatory (i.e.~neutrally stable) solution as long as $-\mu\leq
p\leq\mu$. Note that the stability of the isocurvature modes is
guaranteed even when the fluid pressure vanishes.

Incorporating ${\rm D}_a\omega^2$ in Eq.~(\ref{ilcZdot1}) has
increased the overall resistance of the model against the
gravitational pull of the matter, since gradients in the
rotational energy act like pressure gradients. Their presence is
responsible for the stability of the linear isocurvature
perturbations demonstrated in (\ref{hilDeltaddot}).

\subsubsection{Perturbations under rigid rotation}
Setting ${\rm D}_{a}\omega^{2}=0$ in Eq.~(\ref{lcZdot1}) and using
thee barotropic fluid relation ${\rm D}_ap=(\mu c_{{\rm
s}}^2/a){\cal D}_a$, transforms the system (\ref{lcDdot1}),
(\ref{lcZdot1}) into
\begin{eqnarray}
\dot{{\cal D}}_{a}&=&\omega_{ab}{\cal D}^{b}-(1+w){\cal Z}_{a}\,,
\label{rrlcDdot1}\\ \dot{{\cal Z}}_{a}&=&\omega_{ab}{\cal Z}^{b}-
{\textstyle{\frac{1}{2}}}\kappa\mu(1+3c_{{\rm s}}^{2}){\cal D}_{a}
-\frac{c_{{\rm s}}^{2}}{1+w}{\rm D}_{a}({\rm D}^{b}{\cal
D}_{b})\,.  \label{rrlcZdot1}
\end{eqnarray}
Multiplying the above with the characteristic background scalar
$a$ and then taking their scalar parts, as before, we obtain
\begin{eqnarray}
\dot{\Delta}&=&-(1+w){\cal Z}\,,  \label{rrlDeltadot}\\
\dot{{\cal Z}}&=&-{\textstyle{\frac{1}{2}}}\kappa\mu(1+3c_{{\rm
s}}^{2})\Delta- \frac{c_{{\rm s}}^{2}}{1+w}{\rm D}^{2}\Delta\,,
\label{rrlZetadot}
\end{eqnarray}
which leads to the following wavelike equation for $\Delta_{(k)}$
\begin{equation}
\ddot{\Delta}_{(k)}=\left[{\textstyle{\frac{1}{2}}}
\kappa\mu(1+w)(1+3c_{{\rm s}}^{2})-\frac{k^{2}c_{{\rm
s}}^{2}}{a^{2}}\right]\Delta_{(k)}\,,  \label{hrrlDeltaddot}
\end{equation}
on using the harmonic decomposition
$\Delta=\sum_{k}\Delta_{(k)}Q^{(k)}$ of the previous section. For
dust (i.e.~$w=0=c_{{\rm s}}^{2}$) there is no pressure support and
$\Delta$ grows unimpeded. In the presence of pressure, however,
there is a effective Jeans length given by
\begin{equation}
\lambda_{J}=\frac{c_{{\rm s}}}{\omega\sqrt{1+3c_{{\rm s}}^{2}}}\,,
\label{Jl}
\end{equation}
since $\kappa\mu(1+w)=2\omega^{2}$ to zero order. Inhomogeneities
on scales exceeding $\lambda_{J}$ collapse under the gravitational
pull of the matter, but short wavelength perturbations oscillate
like sound waves. Interestingly, the Jeans scale given above is
comparable to $R_{G}$, namely to the radius of the smallest closed
timelike curves (see~Eq.~(\ref{tlcond})). This means that the
causal regions of a G\"{o}del universe containing a fluid with
non-zero pressure are stable with respect to linear matter
aggregations.

By setting ${\rm D}_{a}\omega^{2}=0$ we have removed the
supporting effect of the centrifugal energy gradients from
Eq.~(\ref{lcZdot1}). This is the reason for the instability
pattern associated with (\ref{hrrlDeltaddot}). The only effect of
the G\"{o}del background is via the rotational energy contribution
to the overall gravitational pull (see also Sec.~3.2). As a
result, the above given Jeans scale is smaller than that of the
almost-FRW models~\cite{EHB}. Note that the Jeans criterion found
here is analogous to that obtained in~\cite{S2}.

\subsection{Directional effects on density perturbations}
The rotation of the G\"{o}del universe introduces a preferred
direction because of the background vorticity vector. The analysis
of density perturbations given in the previous sections examines
the evolution of the scalar $\Delta $, which describes the average
matter aggregation and therefore it does not pick out any
directional effects. These, however, are incorporated in ${\cal
D}_{a}$, the projected vector that describes variations in the
density distribution as seen between two neighbouring observers.
To reveal the directional effects of the background rotation it
helps to implement an additional splitting of the G\"{o}del space,
along and orthogonal to the vorticity vector. As a first step, we
introduce the unit vector $n_{a}=\omega_{a}/\omega$ (where
$\omega=\sqrt{\omega_{a}\omega^{a}}$) parallel to $\omega_{a}$ and
use it to define the 2-dimensional projection tensor
\begin{equation}
f_{ab}=h_{ab}-n_{a}n_{b}\,,  \label{fab}
\end{equation}
with $f_{ab}=f_{(ab)}$, $f_{ab}n^{b}=0$,
$f_{a}{}^{b}f_{b}{}^{c}=f_{a}{}^{c}$ and $f_{a}{}^{a}=2$. This
projects into the 2-D space orthogonal to $\omega _{a}$. Employing
(\ref{fab}) we define the gradient $\tilde{{\rm
D}}_{a}=f_{a}{}^{b}{\rm D}_{b}$, with $n^{a}\tilde{{\rm
D}}_{a}=0$, and introduce the decomposition ${\rm
D}_{a}=\tilde{{\rm D}}_{a}+n_{a}n^{b}{\rm D}_{b}$ orthogonal to
and along the rotation axis. Then, the density gradients split as
\begin{equation}
{\cal D}_{a}=\tilde{{\cal D}}_{a}+\tilde{{\cal D}}n_{a}\,,
\label{cDa2split}
\end{equation}
where $\tilde{{\cal D}}_{a}=f_{a}{}^{b}{\cal D}_{b}=
(a/\mu)\tilde{{\rm D}}_{a}\mu$ is the density perturbation
orthogonal to $n_{a}$ and $\tilde{{\cal D}}=n^{a}{\cal D}_{a}=
(a/\mu )n^{a}{\rm D}_{a}\mu$ is the density perturbation parallel
to the rotation axis. Similarly we write ${\cal Z}_{a}=
\tilde{{\cal Z}}_{a}+\tilde{{\cal Z}}n_{a}$, with $\tilde{{\cal
Z}}_{a}=a\tilde{{\rm D}}_{a}\Theta$ and $\tilde{{\cal Z}}=
an^{a}{\rm D}_{a}\Theta$, for the expansion gradients.

Starting from Eqs.~(\ref{lcDdot1}) and (\ref{lcZdot1}), using the
barotropic fluid relation ${\rm D}_{a}p=c_{{\rm s}}^{2}{\rm
D}_{a}\mu $, expressing ${\rm D}_{a}({\rm D}^{2}p)$ with respect
to ${\rm D}^{2}({\rm D}_{a}p)$ (by changing the order of the
covariant derivatives), and then applying the decomposition
introduced above, we obtain
\begin{equation}
{\ddot{\tilde{{\cal D}}}}_{a}=2(1-2c_{{\rm
s}}^{2})\epsilon_{abc}\omega^{b}{\dot{\tilde{{\cal D}}}}{}^{c}+
{\textstyle{\frac{1}{2}}}\kappa\mu(1+w)(2+c_{{\rm
s}}^{2})\tilde{{\cal D}}_{a}+ c_{{\rm s}}^{2}{\rm
D}^{2}\tilde{{\cal D}}_{a}- 2(1+w)a\tilde{{\rm
D}}_{a}\omega^{2}\,,  \label{cDaorth}
\end{equation}%
for density perturbations orthogonal to $n_{a}$ and
\begin{equation}
{\ddot{\tilde{{\cal D}}}}=
{\textstyle{\frac{1}{2}}}\kappa\mu(1+w)(1+3c_{{\rm
s}}^{2})\tilde{{\cal D}}+ c_{{\rm s}}^{2}{\rm D}^{2}\tilde{{\cal
D}}- 2(1+w)an^{a}{\rm D}_{a}\omega ^{2}\,,  \label{cDapar}
\end{equation}%
for those along the rotation axis. This reveals a qualitatively
different evolution for perturbations orthogonal and parallel to
$\omega_{a}$. The gravitational pull of the matter gets stronger
along the rotation axis as the fluid sound speed increases above
the $c_{{\rm s}}^2=1/2$ threshold. The role of the background
vorticity, on the other hand, is more pronounced orthogonal to the
rotation axis and depends on the nature of the medium and on the
vector product $\epsilon_{abc}\omega^{b}{\dot{\tilde{{\cal
D}}}}{}^{c}$. Note that the effects of pressure gradients and of
gradients in the centrifugal energy density, orthogonal and
parallel to $\omega_{a}$, are effectively identical.

One can obtain more quantitative results by looking at certain
particular cases. For example, consider inhomogeneities parallel
to the rotation axis and assume that the vorticity gradients do
not change along this direction (i.e.~set $n^{a}{\rm
D}_{a}\omega^{2}=0$). Then, it is easy to show that $\tilde{{\cal
D}}$ is unstable for dust and that there is an associated Jeans
length when the fluid has nonzero pressure, as predicted
in~\cite{S2}. Alternatively, we may allow for $n^{a}{\rm
D}_{a}\omega^{2}\neq0$ and assume isocurvature perturbations
(i.e.~set $a{\rm D}_{a}\omega^{2}=\kappa\mu{\cal D}_{a}$). In that
case, inhomogeneities parallel to the rotation axis are neutrally
stable, they oscillate, for all types of matter with $w,\,c_{{\rm
s}}^{2}\geq0$. Thus, the evolution of density perturbations in the
direction of the rotation axis is identical to that of average
matter aggregations.

The supporting role of the gradient ${\rm D}_{a}\omega^{2}$
against inhomogeneities orthogonal to the rotation axis is also
clear, at least for perturbations of the isocurvature type. In
this direction the background vorticity has an additional effect
conveyed by the first term in the right-hand side of
Eq.~(\ref{cDaorth}). This is harder to quantify, however, as it
depends in an intricate way on the nature of the fluid and on the
orientation of the vector
$\epsilon_{abc}\omega^{b}{\dot{\tilde{{\cal D}}}}{}^{c}$.
Interestingly, the effect of this term is reversed as the sound
speed of the medium crosses the $c_{{\rm s}}^2=1/2$
threshold.\footnote{The projected vector $\tilde{{\rm D}}_{a}$
contains information about scalar matter aggregations as well as
for turbulence and distortions in the density distribution of the
fluid. One should be able to isolate and extract this information
by developing further the decomposition introduced via (\ref{fab})
and (\ref{cDa2split}). This, however, goes beyond the scope of the
present paper.}

\subsection{Vorticity perturbations}
Rotation is controlled by a pair of nonlinear propagation and
constraint equations given respectively by
\begin{equation}
\dot{\omega}_{\langle a\rangle}=
-{\textstyle{\frac{2}{3}}}\Theta\omega_{a}-
{\textstyle{\frac{1}{2}}}{\rm curl}\dot{u}_{a}+
\sigma_{ab}\omega^{b}\,,  \label{omegadot}
\end{equation}
and
\begin{equation}
{\rm D}_{a}\omega^{a}=\dot{u}_{a}\omega^{a}\,.  \label{omegacon}
\end{equation}
The same expressions also hold when we linearise about the
G\"{o}del background, although then only the zero-order vorticity
vector contributes to the right-hand side of (\ref{omegadot}) and
(\ref{omegacon}). For a barotropic fluid the vorticity propagation
formula takes the linearised form
\begin{equation}
\dot{\omega}_{\langle a\rangle}=
-{\textstyle{\frac{2}{3}}}\left(1-{\textstyle{\frac{3}{2}}}c_{{\rm
s}}^{2}\right)\Theta\omega_{a}+\sigma_{ab}\omega^{b}\,.
\label{omegadot1}
\end{equation}
When deriving the above we have used the barotropic expression
$\dot{u}_{a}=-[c_{{\rm s}}^{2}/(\mu +p)]{\rm D}_{a}\mu $ for the
4-acceleration, the conservation law (\ref{edc}), and taken into
account that on a rotating background the projected gradients of
scalars do not commute. All these guarantee that ${\rm
curl}\dot{u}_{a}=-2c_{{\rm s}}^{2}\Theta\omega_{a}$.

Part of the rotational behaviour seen in Eq.~(\ref{omegadot1}) is
already familiar from studies of the perturbed FRW universes. In
particular, we see that expansion leads to less rotation when
$c_{{\rm s}}^{2}<2/3$, but increases it as $\omega_{a}$ $\propto
a^{3c_{s}^{2}-2\text{ }}$in models with a stiffer equation of
state for the matter. In the case of contraction the situation is,
obviously, reversed -- at least until the non-linear terms cease
to be negligible. The background rotation of the G\"{o}del
universe, however, has introduced additional effects which
propagate through the shear term in the right-hand side of
(\ref{omegadot1}). To get some idea of impact of the shear, we
will assume that $\omega_{a}$ is a shear eigenvector (i.e.~set
$\sigma_{ab}\omega^{b}=\sigma\omega_{a}$, where $\sigma$ is the
associated eigenvalue). Then, Eq.~(\ref{omegadot1}) simplifies to
\begin{equation}
\dot{\omega}_{\langle a\rangle }= -\left[
{\textstyle{\frac{2}{3}}}\left(1-{\textstyle{\frac{3}{2}}}c_{{\rm
s}}^{2}\right)\Theta-\sigma\right]\omega_{a}\,.  \label{omegadot2}
\end{equation}
Accordingly, when $\sigma>0$ the extra shear coupling will
increase the overall rotation of the model, but it will lead to a
reduction otherwise.

To explain the shear effect, recall that positive $\sigma$ means
that there is an extra expansion along the rotation axis due to
the shear effects. Given, the trace-free nature $\sigma_{ab}$,
this means that there is an overall contraction orthogonal to
$\omega_{a}$, which explains why rotation increases when
$\sigma>0$. Clearly, the opposite happens when $\sigma $ is
negative.

\subsection{Gravitational-wave perturbations}
\subsubsection{Evolution of the shear}
In addition to the vorticity, the shear also affects the linear
evolution of gravitational-wave perturbations. Here we will only
present the relevant propagation and constraint equations, which
will be used to analyse linear gravity waves within a perturbed
G\"{o}del universe in the following sections.

For perfect fluid matter, the nonlinear shear evolution is
governed by a set of one propagation equation and one constraint
\begin{eqnarray}
\dot{\sigma}_{\langle ab\rangle}&=&
-{\textstyle{\frac{2}{3}}}\Theta\sigma_{ab}- E_{ab}+ {\rm
D}_{\langle a}\dot{u}_{b\rangle}- \sigma_{c\langle
a}\sigma_{b\rangle}{}^c- \omega_{\langle a}\omega_{b\rangle}+
\dot{u}_{\langle a}\dot{u}_{b\rangle}\,, \label{sheardot}\\ {\rm
D}^b\sigma_{ab}&=&{\textstyle{\frac{2}{3}}}{\rm D}_a\Theta+ {\rm
curl}\omega_a- 2\epsilon_{abc}\omega^b\dot{u}^c\,.
\label{shearcon}
\end{eqnarray}
When linearised about the G\"odel background, the constraint
equation (\ref{shearcon}) remains the same, whereas the
propagation equation (\ref{sheardot}) reduces to
\begin{equation}
\dot{\sigma}_{ab}=-E_{ab}+ {\rm D}_{\langle a}\dot{u}_{b\rangle}-
\omega_{\langle a}\omega_{b\rangle}\,,  \label{lsheardot}
\end{equation}

\subsubsection{Evolution of the Weyl components}
A covariant description of the gravitational waves is provided by
the electric ($E_{ab}$) and the magnetic ($H_{ab}$) parts of the
Weyl tensor. The two Weyl components support the different
polarisation states of propagating gravitational radiation and
obey evolution and constraint equations that are remarkably
similar to Maxwell's equations. For a perfect fluid, the nonlinear
evolution of the two Weyl components is determined by a set of two
propagation equations~\cite{EvE}
\begin{eqnarray}
\dot{E}_{\langle ab\rangle }&=&-\Theta E_{ab}+{\rm curl}H_{ab}-
{\textstyle{\frac{1}{2}}}\kappa(\mu+p)\sigma_{ab}+
2\dot{u}^{c}\epsilon_{cd\langle a}H_{b\rangle}{}^{d}+
3\sigma_{c\langle a}E_{b\rangle}{}^{c}-
\omega^{c}\epsilon_{cd\langle a}E_{b\rangle}{}^{d}\,,
\label{EWeyldot}\\ \dot{H}_{\langle ab\rangle}&=& -\Theta H_{ab}-
{\rm curl}E_{ab}+ 3\sigma_{c\langle a}H_{b\rangle}{}^{c}-
\omega^{c}\epsilon_{cd\langle a}H_{b\rangle}{}^{d}-
2\dot{u}^{c}\epsilon_{cd\langle a}E_{b\rangle}{}^{d}\,,
\label{HWeyldot}
\end{eqnarray}
supplemented by the constraints
\begin{eqnarray}
{\rm D}^{b}E_{ab}&=&{\textstyle{\frac{1}{3}}}\kappa{\rm D}_{a}\mu+
\epsilon_{abc}\sigma^{b}{}_{d}H^{cd}- 3H_{ab}\omega^{b}\,,
\label{EWeylcon}\\ {\rm D}^{b}H_{ab}&=&\kappa(\mu+p)\omega_{a}-
\epsilon_{abc}\sigma^{b}{}_{d}E^{cd}+ 3E_{ab}\omega^{b}\,.
\label{HWeylcon}
\end{eqnarray}
In addition, the magnetic Weyl tensor is related to the kinematic
variables through the nonlinear constraint
\begin{equation}
H_{ab}={\rm curl}\sigma_{ab}+ {\rm D}_{\langle
a}\omega_{b\rangle}+ 2\dot{u}_{\langle a}\omega_{b\rangle}\,,
\label{HWeyl}
\end{equation}
where ${\rm curl}T_{ab}=\epsilon_{cd\langle a}{\rm
D}^{c}T^{d}{}_{b\rangle}$ for any symmetric, orthogonally
projected tensor $T_{ab}$.

Linearised about the unperturbed G\"{o}del background, the
propagation equations (\ref{EWeyldot}) and (\ref{HWeyldot}) become
\begin{eqnarray}
\dot{E}_{\langle ab\rangle}&=&-\Theta E_{ab}+ {\rm curl}H_{ab}-
3\omega^{c}\sigma_{c\langle a}\omega_{b\rangle}-
\omega^{c}\epsilon_{cd\langle a}E_{b\rangle}{}^{d}\,,
\label{lEWeyldot}\\ \dot{H}_{ab}&=&-{\rm curl}E_{ab}-
\omega^{c}\epsilon_{cd\langle a}H_{b\rangle}{}^{d}-
2\dot{u}^{c}\omega_{c\langle a}\omega_{b\rangle}\,,
\label{lHWeyldot}
\end{eqnarray}
on using the zero-order relation (\ref{sigmadot}) to express the
background $E_{ab}$ tensor with respect to the vorticity vector.
Similarly, the two constraints (\ref{EWeylcon}) and
(\ref{HWeylcon}) reduce to
\begin{eqnarray}
{\rm D}^{b}E_{ab}&=&{\textstyle{\frac{1}{3}}}\kappa{\rm D}_{a}\mu-
3H_{ab}\omega^{b}\,,  \label{lEWeylcon}\\ {\rm D}^{b}H_{ab}&=&
\kappa(\mu+p)\omega_{a}+
\epsilon_{abc}\sigma^{b}{}_{d}\omega^{c}\omega^{d}+
3E_{ab}\omega^{b}\,.  \label{lHWeylcon}
\end{eqnarray}
while (\ref{HWeyl}) does not change.

\subsubsection{Isolating the gravitational-wave perturbations}
Given the symmetric and trace-free nature of $E_{ab}$ and
$H_{ab}$, we can isolate the pure tensor modes, namely the
gravitational waves, by demanding that $E_{ab}$ and $H_{ab}$ are
also divergence-free at the linear level. In practice, this means
setting the right-hand side of (\ref{lEWeylcon}) and
(\ref{lHWeylcon}) to zero. When there is no background vorticity,
as in perturbed FRW models, we can isolate the gravity waves by
assuming a barotropic fluid (i.e.~setting $p=p(\mu ))$ and by
switching off both scalar and vector perturbations. This is done
by introducing the linear constraints ${\rm D}_{a}\mu=0={\rm
D}_{a}\Theta=\omega_{a}$, which are consistent to first
order~\cite{DBE}-\cite{T}. When the background is rotating,
however, the aforementioned set of constraints is not enough.
Moreover, switching off the model's rotation is no longer an
option. Here, we are dealing with the rotating G\"{o}del
background, and we will isolate the pure tensor modes by imposing
the following self-consistent linear constraints
\begin{eqnarray}
{\rm D}_{a}\mu=0={\rm D}_{a}\Theta\,, &\hspace{15mm}& {\rm
D}_{a}\omega^{2}=0={\rm curl}\omega_{a}\,,  \label{pt12}\\
H_{ab}\omega^{b}=0=\sigma_{ab}\omega^{b}\,, &\hspace{15mm}&
E_{ab}\omega^{b}=-{\textstyle{\frac{1}{3}}}\kappa(\mu+p)\omega_{a}\,,
\label{pt34}
\end{eqnarray}
in addition to the barotropic fluid assumption. Clearly these
constraints guarantee the transversality of $E_{ab}$ and $H_{ab}$,
as well as that of $\sigma_{ab}$, while they maintain a non-zero
background rotation. Note that constraint (\ref{pt12}a) is the
standard restriction imposed on perturbed FRW cosmologies, while
the rest are new. Of the three extra constraints, (\ref{pt12}b)
guarantees that any perturbations in the centrifugal energy that
happen to be present are switched off, and that the vorticity
vector is curl-free. Constraints (\ref{pt34}), on the other hand,
imply that $\omega_{a}$ is an eigenvector of $H_{ab}$,
$\sigma_{ab}$ and $E_{ab}$ to linear order. In the first two cases
the corresponding eigenvalues are zero, whereas in the third the
eigenvalue is $-\kappa(\mu +p)/3$. Finally, we point out that the first
of the constraints (\ref{pt12}a), together with the barotropic
fluid assumption, guarantees that ${\rm D}_{a}p=0$. This in turn
ensures that $\dot{u}_{a}=0$ through the momentum-density
conservation. As a result, ${\rm D}_{a}\omega^{a}=0$ (see
Eq.~(\ref{omegacon})), which means that the vorticity vector is
solenoidal.

We check the consistency of any set of constraints by propagating
them in time. If every constraint is still satisfied, without the
need of any extra restrictions, we say that the set is
self-consistent. Here we can show that the set (\ref{pt12}),
(\ref{pt34}) is self-consistent only when the matter source is
pressure-free dust (i.e.~for $p=0$).\footnote{In principle, there
might be an alternative set of linear constraints, which is less
restrictive than (\ref{pt12}), (\ref{pt34}) and still isolates the
pure tensor modes of a perturbed G\"{o}del universe. It is also
possible that a modified set of constraints could isolate the
linear gravity-wave perturbations, even for matter with nonzero
pressure. As yet, however, we have not been able to identify any
such sets.} In particular, the consistency of (\ref{pt12}a)
follows directly from the linear propagation equations
(\ref{lcDdot1}), (\ref{lcZdot1}) of the density and the expansion
gradients respectively. The consistency of the two constraints in
(\ref{pt12}b) is shown by propagating them in time and then by
using the linear commutation laws between time derivatives and
projected covariant derivatives (see~\cite{vE} for a comprehensive
list). Finally, taking the time derivatives of $H_{ab}\omega^b$,
$\sigma_{ab}\omega^b$ and $E_{ab}\omega^b$ we can show that
constraints (\ref{pt34}) are also consistent to linear order. In
doing so, one needs to employ the zero order relations
(\ref{sigmadot}), (\ref{rho+p}) and use the covariant identities
given in~\cite{vE}.

\subsubsection{Linear evolution of gravitational waves}
Once the constraints (\ref{pt12}), (\ref{pt34}) have been imposed
and the pure tensor modes have been isolated,
Eqs.~(\ref{lEWeyldot}) and (\ref{lHWeyldot}) reduce to
\begin{eqnarray}
\dot{E}_{ab}&=&-\Theta E_{ab}+ {\rm curl}H_{ab}-
\omega^{c}\epsilon_{cd\langle a}E_{b\rangle}{}^{d}\,,
\label{lEWeyldot1}\\ \dot{H}_{ab}&=&-{\rm curl}E_{ab}-
\omega^{c}\epsilon_{cd\langle a}H_{b\rangle}{}^{d},
\label{lHWeyldot1}
\end{eqnarray}
respectively. In the same situation, the constraint (\ref{HWeyl})
reads
\begin{equation}
H_{ab}={\rm curl}\sigma_{ab}+ {\rm D}_{\langle
a}\omega_{b\rangle}\,, \label{HWeyl1}
\end{equation}
implying that $H_{ab}\neq{\rm curl}\sigma_{ab}$ in contrast to the
perturbed FRW case. In the G\"{o}del spacetime the magnetic
component of the Weyl tensor vanishes identically (this is a
feature that permits the existence of the close Newtonian analogue
discussed above). Its electric counterpart, however, has a nonzero
value that is expressed in terms of the model's vorticity (see
Eq.~(\ref{sigmadot})). This implies, according to~\cite{SW}, that
only the gauge-invariance of $H_{ab}$ is guaranteed when
linearising about a G\"{o}del background. Therefore, to avoid any
gauge-related ambiguities, we will monitor the linear gravity-wave
perturbations by looking exclusively into the evolution of
$H_{ab}$.

To obtain the wave equation for $H_{ab}$ we take the time
derivative of Eq.~(\ref{lHWeyldot1}) and then employ a lengthy,
but relatively straightforward, calculation. In the process we
apply the constraints (\ref{pt12}), (\ref{pt34}), use the zero
order relations (\ref{sigmadot}) and (\ref{rho+p}), to express
$E_{ab}$ and $\kappa(\mu+p)$ with respect to the background
vorticity vector, and employ the commutation laws and the
covariant identities of~\cite{vE}. At the end we arrive at
\begin{equation}
\ddot{H}_{ab}-{\rm D}^{2}H_{ab}=0\,,  \label{ddotHWeyl}
\end{equation}
which means that the linear order wave equation of $H_{ab}$ has no
vorticity-related source terms, despite the fact that the
background rotation has not been switched off. Following the
above, we claim that the G\"{o}del universe is neutrally stable
against these linearised gravity-wave perturbations.

\section{The case of homogeneous perturbations}
We can also consider the stability of the G\"{o}del universe
against homogeneous perturbations, by ignoring spatial gradients
such as ${\rm D}_{a}\mu$, ${\rm D}_{a}\Theta$, ${\rm
D}_{a}\omega^{2}$, etc. We will do this by focusing on the
Raychaudhuri equation, which takes the linearised form
\begin{equation}
\dot{\Theta}=-{\textstyle{\frac{1}{2}}}(\mu+3p)+ \omega^{2}+
\Lambda\,.  \label{homRay}
\end{equation}
We can determine the effects of small homogeneous perturbations of
the metric that alter the matter density and rotation by comparing
the relative growth of the vorticity and matter energy-density
terms in (\ref{homRay}). The conservation of angular momentum for
rotational perturbations of a perfect fluid with equation of state
\[
p=(\gamma-1)\mu\equiv c_{s}^{2}\mu
\]
requires that the angular momentum of an eddy of mass $M$ and
comoving scale $a$ be constant; thus
\[
Ma^{2}\omega=const.
\]
Since $\mu\propto a^{-3\gamma}$ and $M\propto\mu a^{3\text{ }}$we
have $\omega\propto a^{3\gamma-5}$ (in agreement with Eq.
(\ref{omegadot2}) when $\sigma $ is negligible) and so
\[
\frac{\omega^{2}}{\mu}\propto a^{9\gamma-10}.
\]
Thus, for changes that cause expansion ($a\rightarrow\infty$) the
rotation dominates the effects of self gravity whenever the fluid
state satisfies $\gamma>10/9$. This includes the case of radiation
($\gamma=4/3$) but excludes that of dust ($\gamma=1$). For
perturbations that produce gravitational collapse
($a\rightarrow0$) the opposite conclusion holds: rotation becomes
negligible with respect to the self gravity of the fluid whenever
$\gamma>10/9$ but is dominant whenever $\gamma<10/9.$
Superficially, this implies that rotation would halt gravitational
collapse whenever $\gamma<10/9,$ but the self-gravitating effect
of the rotation actually contributes to the collapse and a
singularity will result so long as the matter stresses obey the
strong energy condition, in accord with the singularity
theorems~\cite{HE}. The detailed behaviour of the rotational
collapse depends on the non-linear behaviour of the perturbations
as $a\rightarrow0$.

\section{Scalar fields in the G\"odel universe}
Scalar field dominated models are key to contemporary cosmology,
given their prominent role in inflationary scenarios for the very
early universe and as candidates for the dark energy content of
the universe today. It is interesting to examine whether the high
symmetry of homogeneous spacetimes might make them more likely as
initial states. If so, they may constraint the existence of scalar
fields in some way. The symmetries of the Einstein static
universe, for example, mean that any scalar field that happened to
be present will have constant kinetic energy and
potential~\cite{BEMT}. Here we will look into the implications of
the G\"{o}del symmetries for such a scalar field

Consider a general spacetime filled with a minimally-coupled
scalar filed $\phi$ and assume that
$\nabla_{a}\phi\nabla^{a}\phi<0$~\cite{BED}. Then, $\phi$ has a
perfect fluid description with respect to a 4-velocity field
$u_{a}$ that is normal to the surfaces $\{\phi=$ constant$\}$.
More specifically, the scalar field behaves as a perfect fluid
with
\begin{equation}
\mu _{\phi }={\textstyle{\frac{1}{2}}}\dot{\phi}^{2}+ V(\phi)
\hspace{10mm} {\rm and} \hspace{10mm} p_{\phi}=
{\textstyle{\frac{1}{2}}}\dot{\phi}^{2}- V(\phi)\,, \label{sfmup}
\end{equation}
relative to $u_{a}=-\nabla_{a}\phi/\dot{\phi}$~\cite{BED}. Note
that $\dot{\phi}=-(\nabla_{a}\phi\nabla^{a}\phi)^{1/2}$ determines
the kinetic energy of the scalar field and $V(\phi)$ is the
associated potential.

The perfect fluid description of $\phi$ is necessary if one wants
to allow scalar fields in the G\"{o}del universe. However, since
the covariant derivatives of scalars commute, it becomes clear
that the 4-velocity field $u_{a}$ defined above is irrotational.
In other words, the vorticity vector associated with $u_{a}$
vanishes~\cite{BED}, which is in direct contradiction with the
generic rotation of the G\"{o}del spacetime. Putting it in a
different way, one can say that the presence of the above $u_{a}$
implies the existence of global spacelike hypersurfaces, something
strictly forbidden in the G\"{o}del spacetime. On these grounds,
one could argue that, at least globally, minimally-coupled scalar
fields are not compatible with the symmetries of the G\"{o}del
universe. More generally, these considerations may offer some
insight into the absence of observational evidence for rotation in
the universe today, to very high precision \cite{BJS}. If the
initial state of the universe is dominated by scalar fields, which
appear to exist in profusion in string theories \cite{suss}, then
a zero-vorticity initial state will be enforced and provide a
simple explanation for Mach's 'Principle' even without inflation.
Of course, any subsequent bout of inflation driven by scalar
fields will reduce the effects of pre-existing vorticity to levels
far below the threshold of detectability today and would be unable
to generate new vorticity from the scalar fluctuations. As a
result, the observation of any large-scale vorticity in the
universe would be a decisive piece of evidence against an early
inflationary state \cite{BL} and reveal specific information about
the nature of matter at very high energies.

It should be emphasised that in this section we have considered an
exact G\"{o}del spacetime containing a single scalar field with a
timelike gradient (i.e.~$\nabla_a\phi\nabla^a\phi<0$) and a
corresponding Segr\`{e} type $[1,(111)]$. However, scalar fields
of different nature are not a priori excluded. For example, a
scalar field with zero gradient (i.e.~$\nabla_a\phi=0$), which
corresponds to an effective cosmological constant and has a
Segr\`{e} type $[(1,111)]$ is clearly compatible with the
symmetries of the G\"{o}del metric. For more details on the
classification of scalar fields with respect to the nature of
their gradients the reader is referred to~\cite{SRT}. We should
also point out that scalar fields are allowed in rotating
G\"{o}del-type spacetimes as it has been shown in~\cite{RT}.

\section{Discussion}
G\"{o}del's universe is one of the most intriguing solutions of
the Einstein field equations. The common factor behind its unique
and exotic features is the rigid rotation of the model, which is
why the G\"{o}del solution has been widely used to illustrate the
possible general-relativistic effects of global vorticity and
time-travel. In the present article we have looked into the
implications of the model's generic rotation for the stability of
the G\"{o}del universe under a variety of perturbations.

We first considered scalar-matter aggregations and looked into two
different types of inhomogeneities, namely isocurvature and
perturbations under rigid rotation. We found different patterns of
stability determined by the presence or absence of gradients in
the centrifugal energy. These act as effective pressure gradients
balancing the gravitational pull of the matter fields.
Interestingly, the latter is found to have a contribution from the
rotational energy as well. This is a purely relativistic effect,
as opposed to the supporting effect of the vortical energy
gradients which is Newtonian in nature. When the gradients in the
rotational energy are included, as occurs for the isocurvature
perturbations, the model is found to be stable against density
inhomogeneities even for pressure-free fluids. In their absence,
however, stability is possible only if there are pressure
gradients and then only on scales below an effective Jeans length.
The latter is of the order of the largest causal region and
smaller than its counterpart in almost-FRW cosmologies. This is
what happens when dealing with perturbations under rigid rotation.

The equation of state of the cosmic medium is decisive for the
evolution of homogenous perturbations. These were studied by
comparing the relative growth between the vorticity and the matter
terms in the Raychaudhuri equation. For perturbations causing
expansion, we found that rotation dominates over self-gravity when
the equation of state obeys $p>\mu/9$, which includes the cases of
radiation and stiff matter. In the case of contraction the
situation is reversed.

The linear rotational behaviour of the perturbed G\"{o}del
universe was found to follow the general pattern familiar from the
almost-FRW studies, with some additional effects due to shape
distortions. We examined these effects by aligning the vorticity
vector along the shear eigenvectors. We found that, when the shape
distortions lead to an extra contraction orthogonal to the
rotation axis, the vorticity of the model increased, but it
decreased otherwise.

The covariant analysis of linear gravitational waves on rotating
backgrounds is complicated by the need to select and impose a set
a constraints that isolates the pure tensor modes without
switching off the vorticity. Here, we have introduced for the
first time, a set of seven linear constraints that does this. We
then proceeded to study the gravity-wave evolution by looking at
the behaviour of the magnetic Weyl tensor. Our choice was based on
the fact that this tensor has no Newtonian analogue and vanishes
in the exact G\"{o}del spacetime. The latter property guarantees
that our analysis is free of any gauge-related ambiguities. The
result was a simple plane-wave equation for $H_{ab}$ without any
vorticity-related source terms, which argues for the neutral
stability of linear gravity-wave perturbations.

Finally, we have considered the implications of the G\"{o}del
symmetries for the presence of scalar field. We argue that the
absence of any global spacelike hypersurfaces in the G\"{o}del
spacetime, namely the absence of any global irrotational vector
field, forbids the introduction of scalar fields with timelike
gradient vectors.

\section{Acknowledgements}
The authors would like to thank Anthony Challinor, Alan Coley,
Kostas Kokkotas, Jeff Murugan, Nikos Stergioulas, Henk van Elst
and particularly George Ellis for many helpful discussions and
comments. CT was supported by a Sida/NRF grant, a DAA bursary and
PPARC and wishes to thank the Centre for Mathematical Sciences at
DAMTP, where part of this work was done, for their hospitality.

\end{document}